\documentclass[letter,preprint]{ptptex}
\usepackage{graphicx}

\notypesetlogo 
\preprintnumber[3.5cm]{
UT-Komaba/11-1v3}

\title{Deciphering the Measured Ratios of Iodine-131 to \\Cesium-137  
at the Fukushima Reactors}

\author{T. \textsc{Matsui}}

\inst{Institute of Physics, University of Tokyo, Komaba, Tokyo 153-8902, Japan}

\abst{We calculate the relative abundance of the radioactive 
isotopes Iodine-131 and Cesium-137 produced by nuclear fission in 
reactors and compare it with data taken at the troubled Fukushima 
Dai-ichi nuclear power plant.    
The ratio of radioactivities of these two isotopes can be used to obtain   
information about when the nuclear reactions terminated. 
}

\begin{document}
\maketitle

\vspace{5mm}
\noindent
1. {\it Introduction}\hspace*{7mm}  
The current state of the troubled Fukushima Dai-ichi nuclear 
power plant is of considerable concern.  The plant,
on the east shore of Fukushima prefecture was first 
hit, at 14:46 pm local time on March 11, 2011, by  
the magnitude 9.0 earthquake
on the Richter scale about 
150 km east of Honshu, the main island of Japan, and then hit
by a giant tsunami, exceeding 14~m high, about one hour after the earthquake. 
Although three reactors (Units 1, 2, 3), which had been on operation at the plant, were 
automatically shut down immediately after the earthquake, the emergent electric power 
generators, working after the loss of external power supply, 
for cooling the reactor system were knocked out by the tsunami.
Uncontrolled decay heat from the fuel rods has caused serious trouble in 
containing the radioactive materials which had been accumulated in the reactors and 
in the spent-fuel rods in the cooling pools.   
Explosions of hydrogen gas, created by the reaction of the heated Zirconium fuel 
cladding with water vapor, took place at the buildings of the Units 1, 3, and 4 reactors, 
severely damaging those buildings. 

At the moment, only very limited information is available about the interior
conditions of the reactor complex owing to the highly radioactive environment, with
intense gamma rays from radioactive materials that apparently escaped from the reactors
preventing direct human access for inspections. 
However, certain information has been released by the Tokyo Electric 
Power Company (TEPCO) \cite{TEPCO} on measurements of the radioactivity at several 
monitoring posts near the troubled reactors and from a water sample taken from 
the spent-fuel cooling pool of the Unit-4 reactor building.\cite{site4}

We call attention here to the significance of the relative strength of the reported radioactivities 
of different radioactive isotopes.   In particular, we focus on the two radioactive isotopes 
abundantly produced by fission, the Iodine isotope I-131 and the Cesium isotope Cs-137; 
the former has the half-life of 8 days while for the latter it is 30 years.   
The ratio should decrease on the time scale of days after the termination of 
nuclear fission processes and hence may be used for measuring the age of 
the fission products, similar to the carbon dating method using 
the ratio of C-14 to C-12 
in the remains of ancient life\cite{carbon}. 

One crucial difference between the proposed iodine-cesium dating and 
 carbon dating is that  while the initial condition of the latter is essentially 
fixed by the equilibrium C-14/C-12
ratio in atmosphere with respect to
cosmic ray interactions, the initial condition of the former depends on the 
prehistory of controlled nuclear reactions in the reactor.
Another potential problem, which does not exist for carbon dating, is 
the effect of the different chemical properties of iodine and cesium\footnote{This problem can be avoided by using Cs-134 and Cs-137.  
However, Cs-134 has a half-life of 2 years which is too long to see observable effects 
in the currently available data.}   
possibly reflected in the change of their relative abundance depending on 
the environment in which they are found, for example in the water (fresh or sea),
in the air, or in aerosols\cite{chemistry}.    

With these caveats in mind, we make a simple estimate of time-dependent numbers of 
I-131 and Cs-137 nuclei produced from a uranium burning nuclear reactor and then examine
the data of the water samples released by TEPCO taken at several monitoring posts near 
the troubled Fukushima nuclear reactor plant, including sea water at the Southern Discharge 
Channel, Unit-4 cooling pool, Sub-drain (under ground) water, and sea water taken 
at the Intake Canal.   Some of these data show anomalously high iodine-cesium ratios
compared to the values expected for the nuclear fuels in the reactors if the nuclear fission 
has indeed terminated before the outbreak of the accident.   We draw tentative conclusions 
from these analyses of the data at the end.\\

\noindent
2. {\it Iodine-cesium ratios in nuclear reactors}\hspace{7mm}
We assume that in an operating nuclear reactor, where criticality is 
sustained for chain nuclear fission reactions, these isotopes are produced with 
a constant production rate, while the produced isotopes decay exponentially 
with a constant decay rate.  
The number of I-131 isotopes created by a nuclear reactor which had been in operation 
from time $t_i$ to $t_f$ is then given at later time $t$ by\footnote{This 
result may be obtained by integration of the equation, 
$ d N_{\rm I}/dt  = f_{\rm I}  {\cal N}_0 \theta (t; t_1, t_f ) - \lambda_{\rm I} N_{\rm I}$, where
$\theta (t; t_i, t_f ) = 1$ for $t_i < t < t_f$ and $\theta (t; t_i, t_f ) = 0$ otherwise, with the
initial condition $N_{\rm I} (t_i) = 0$.}
\begin{equation}
N_{\rm I} (t) = \int_{t_i}^{t_f} d t' f_{\rm I} {\cal N}_0 e^{-\lambda_{\rm I} (t - t') },
\end{equation}
where ${\cal N}_0$ is the number of fissions per unit time, $f_{\rm I}$ is the fraction 
of the I-131 produced per fission, and the decay
rate $\lambda_{\rm I}$ is given in terms of the half-life by 
$\tau_{\rm I} = \ln 2/ \lambda_{\rm I} =$ 8 days.
The integration gives 
\begin{eqnarray}
N_{\rm I} (t) & = & f_{\rm I} {\cal N}_0 e^{-\lambda_{\rm I} (t - t_f)} 
\frac{1 - e^{-\lambda_I (t_f - t_i)} }{\lambda_{\rm I}}  \nonumber \\
& \simeq &  \frac{f_{\rm I} {\cal N}_0}{\lambda_{\rm I}}  e^{-\lambda_{\rm I} (t - t_f) },
\end{eqnarray}
where we have assumed that 
the working time of the reactor is much longer than the half-life of I-131, $\lambda_{\rm I} (t_f - t_i) \gg 1$.
The abundance of I-131 saturates on a time scale $\tau_I$ and
remains constant during the operation of the reactor.
 
A similar estimate can be made for Cs-137.  In this case, however, the decay 
of Cs-137 is very slow, with a long half-life 
$\tau_{\rm Cs} = \ln 2/ \lambda_{\rm Cs} =$ 30 years; hence 
$\lambda_{\rm Cs} (t_f - t_i) \ll 1$ and we obtain 
\begin{eqnarray}
N_{\rm Cs} (t) & = & f_{\rm Cs} {\cal N}_0 e^{-\lambda_{\rm Cs} (t - t_f) } 
\frac{1 - e^{-\lambda_{\rm Cs} (t_f - t_i)} }{\lambda_{\rm Cs}} \nonumber  \\
& \simeq &  f_{\rm Cs} {\cal N}_0 \Delta t  e^{-\lambda_{\rm Cs} (t - t_f) }, 
\end{eqnarray}
where $\Delta t =  t_f - t_i$ is the time interval that the reactor has been in operation. 
The Cs-137 accumulates during the operation of the 
reactor due to its long lifetime, and hence becomes proportional to $\Delta t$.

The radioactivity of each of these elements measured in becquerel (Bq) 
is given by the decay rates 
\begin{equation}
\Gamma_{\rm I} (t ) = - \frac{d N_{\rm I}}{dt } =  f_{\rm I} {\cal N}_0  e^{-\lambda_{\rm I} (t - t_f) } 
\end{equation}
and
\begin{equation}
\Gamma_{\rm Cs} (t) = - \frac{d N_{\rm Cs}}{dt } =  f_{\rm Cs} {\cal N}_0  \lambda_{\rm Cs} \Delta t 
e^{-\lambda_{\rm Cs} (t - t_f) }, 
\end{equation}
so that the ratio becomes
\begin{equation}
R_{\rm I/Cs} (t) = \frac{f_{\rm I}}{f_{\rm Cs}} \frac{\tau_{\rm Cs} }{ \Delta t \ln 2} \left( \frac{1}{2} 
\right)^{(t - t_f)/\tau_{\rm I}},
\label{R}
\end{equation}
where we have again used the condition $\lambda_{\rm Cs} (t - t_f) \ll 1$.
This formula gives the average ratio of the radioactivities of I-131 and Cs-137 contained 
in the fuel rods after the termination of the nuclear reaction at $t_f$. 

The fractions of each fission product are $f_{\rm I} = 2.88 \times 10^{-2}$,
$f_{\rm Cs} = 6.22 \times 10^{-2}$ for the cumulative thermal fission yields 
of U-235, taken from the IAEA data base\cite{IAEA}.  
For the three Fukushima Dai-ichi nuclear reactors in operation at the time $t_X$ of the
earthquake,  the ratio at $t = t_X$ was
 \begin{equation}
R_{\rm I/Cs} (t_X) = \frac{f_{\rm I}}{f_{\rm Cs}} \frac{\tau_{\rm Cs} }{ \Delta t \ln 2} 
= \frac{240 }{\Delta t},
\end{equation}
where $\Delta t$ is measured in months.  If we choose $\Delta t = 7 (12)$ months
then $ R_{\rm I/Cs} (t_X) = 34 (20)$. 
On April 12, this ratio would be reduced by factor $2^{32/8} = 16$ to 2.1 (1.3). \\

 
\noindent 
3. {\it Deciphering the Fukushima nuclear accident data} \hspace{7mm}
We now examine the data released by TEPCO for the Fukushima nuclear accident 
in the light of our theoretical estimates of the ratios of the Iodine-131 and Cesium-137 
accumulated in the nuclear reactors and attempt to read their implications. 
 
\begin{figure}[t]
\begin{center}
\resizebox*{!}{6.6cm}{\includegraphics{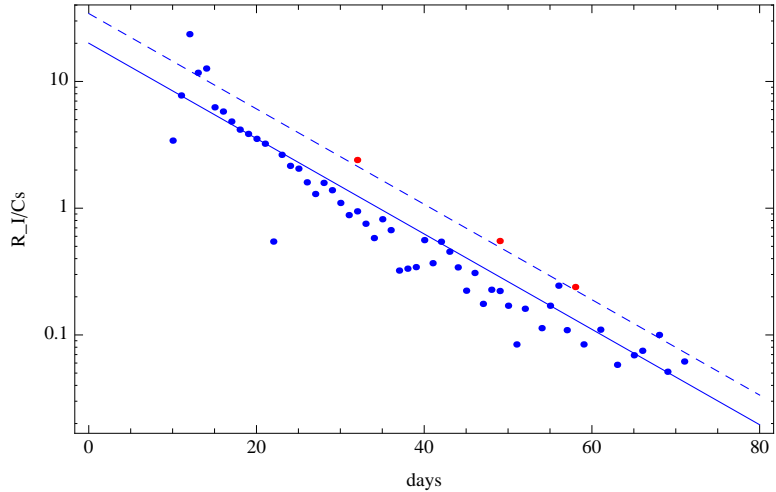}}
\end{center}
\caption{\label{seawater} (color online)
Ratio of the radioactivities of I-131 and Cs-137 plotted against number of days since the 
earthquake
(March 11, 2011). The blue solid line ($\Delta t = 12$ months) and the blue dashed line 
($\Delta t = 7$ months) 
are the theoretical lines based on the formula (\ref{R}); the blue dots are the measured ratios
from the samples of sea water taken at the southern monitoring post.  The red dots are the 
data of the water samples of the Unit-4 reactor cooling pool.
}
\end{figure}

(i) {\it Sea water at the Southern Discharge Channel:}
In Fig.~\ref{seawater}, we show a semi-log plot of the ratio (\ref{R}) against the number of days 
past X-day (March 11, 2011) with two different values for $\Delta t$; 
the upper line corresponds to $\Delta t = 7$ months and the lower line
to $\Delta t =1$ year.  
Note that the slopes of these lines are fixed by the half-life of I-131; only the
normalization changes with a change of $\Delta t$.  
The larger the value of $\Delta t$, the lower the position of the line.  
The blue dots are the data of the sea water samples taken at the South Discharge 
Channel, a monitoring post located 330~m south of the Discharge Channel of  the Unit-1 -- 4 
reactor sites.\cite{seawaterdata}  
The data are better fit by the lower line with $\Delta t = 1$ year, but the slope is 
a little steeper for the data taken for the first 50 days.
As mentioned, the slope of the decay line is essentially fixed by the half-life of
I-131, so this may at first look puzzling. 
The change of the slope may, however, well have been caused by the mixture of 9000 tons 
of old low level contaminated water which had been discharged from the Central 
Radioactive Waste Disposal Facility from April 4 to 10\cite{discharge}.  
In these wastes I-131 had been completely depleted and only long half-life
isotopes like Cs-137 were present. 
The slope tends to decrease slightly later. 
This change may be caused by the leakage of contaminated water with large iodine-cesium ratio 
from Unit-2 reactor site, as we shall see later.

Although we have ignored the effects caused by the chemical or physical differences
of iodine and cesium in compound formation, ionization, diffusion, etc.,
the overall fit of the sea water data with the present theoretical calculation is very 
encouraging. 

(ii) {\it Unit-4 cooling pool: }
We now turn our attention to the spent-fuel rods in the Unit-4 cooling pool. 
Note that the Unit-4 reactor was not in operation at the time of the earthquake. 
The same ratio for the fission products contained in the spent-fuel rods in the Unit-4 cooling 
pool should be smaller by an extra attenuation factor  
\begin{equation}
\left( \frac{1}{2} \right)^{(t_X - t_f)/\tau_{\rm I}} = 1/2^{90/8} \simeq 1/2400,
\end{equation}
if all these fuel rods were intact since they had been removed from the Unit-4 reactor
three months earlier (December 2010).  
Since some of the fuel could be older,  this 
ratio may be even smaller\cite{skimmer}.  
The calculated ratio is not, however, consistent with what has been reported.    
According to the TEPCO press release on April 14\cite{site4}, the radioactivities of 220 
Bq/cc of I-131 and 93 Bq/cc of Cs-137 have been detected from the water sample taken 
on April 12 from the Unit-4 cooling pool, where more than 1500 spent-fuel rods were stored.    
The data were analyzed on April 13.  The ratio 2.4 is closer to the values of the ratio of 
the radioactivities from the other reactors, as shown by the red dot in Fig.~\ref{seawater}.
\cite{newdata}

If the data are correct, they would imply that at least some of the radioactive fission 
products found in the Unit-4 cooling pool have been produced by nuclear reactions
that took place at  around time $t_X$ or later.\footnote{
We have tacitly assumed that the solubility of iodine or cesium 
in water is not changed considerably by boric acid, which has been added to  
the water to suppress chain nuclear reactions.} 
One possible explanation could be contamination by fallout of the fission products 
created in the neighboring reactors.\footnote
{
Actually, the hydrogen explosions of  the Unit-4 reactor building took place on March 15 
after the explosion of the Unit-3 reactor building on March 14. 
However, the explosion on March 15 may have sent fallout from the hydrogen explosion 
of the Unit-3 building on the roof of the Unit-4 reactor building into the pool. 
The author thanks Professor Ichimura for calling his attention to these subtle points.}
The iodine-cesium ratio of the water sample taken from the Unit-3 cooling pool and measured 
on May 8, however,  has turned out to be 0.079\cite{pool3}; this value is about factor 3 smaller 
than the same ratio 0.24 for the water of Unit-4 cooling pool measured on the same day\cite{pool40},
but instead fits to the decay line of the sea water data, suggesting that the radioactivity of  
the sea water has been caused by the fall-out of the radioactive material released by the hydrogen 
explosions. 
Although the difference of the ratios in Unit-3 and Unit-4 cooling pools may be caused
by fluctuations of the distribution of the fallout,  we may take this difference more seriously
and seek other explanation for the origin of high iodine-cesium ratio found in the Unit-4 cooling pool.    
Another possible explanation could be that a nuclear 
chain reaction was reignited in the melted used fuel in the Unit-4 cooling pool for a certain period. 

We now explore whether criticality had been re-established only briefly, in a 
period $\delta t$ from time $t'_i$ to $t'_f = t'_i + \delta t$.  For 
simplicity in the following analysis, we assume $\delta t \ll \tau_{\rm I} = 8$ 
days.   This would be reasonable since the heat produced by a nuclear
reaction in a chunk of the melted fuel would cause rapid disintegration of the
chunk, terminating criticality.   
In this case, the number of newly produced I-131 nuclei increases in  proportion
to $\delta t$:
\begin{eqnarray}
\delta N_{\rm I} (t) 
& \simeq &  f_{\rm I} {\cal N}_1 \delta t  e^{-\lambda_{\rm I} (t - t'_f) }, 
\end{eqnarray}
where ${\cal N}_1 $ is the average number of nuclear fissions per unit time 
taking place during this brief time period.     
Similarly the number of newly produced Cs-137 nuclei is  
\begin{eqnarray}
\delta N_{\rm Cs} (t) 
& \simeq &  f_{\rm Cs} {\cal N}_1 \delta t  e^{-\lambda_{\rm Cs} (t - t'_f) },
\end{eqnarray}
so that the ratio of the radioactivities of the newly produced I-131 and Cs-137 would be 
\begin{equation}
R^{\rm new}_{\rm I/Cs} (t) = \frac{f_{\rm I}}{f_{\rm Cs}} \frac{\tau_{\rm Cs}}{\tau_{\rm I} } 
e^{-\lambda_{\rm I} (t - t'_f) } 
= 630 \times \left( \frac{1}{2} \right)^{(t - t'_f)/\tau_{\rm I}}  .
\label{Rmax}
\end{equation}
This result with the replacement $t'_f \to t_X$ also gives an upper bound on 
the radioactivity ratio for the fission products created in the reactor under 
normal conditions before X-day for short $\Delta t$.  

The radioactivity measured in the sample water taken from the Unit-4 cooling 
pool could perhaps be from a mixture of the newly created radioactive fission 
products and the remnant of old fission products which is mostly composed of 
Cs-137, since I-131 has been depleted considerably, leading to the ratio 2.4 
accidentally coincident with the ratio from the other reactors.  
In any event, the amount of new fission products may be small since 
the absolute value of the measured radioactivity in the sample is not 
significant.\footnote{ The total amount of the radioactivity by I-131 contained in the water of 
Unit-4 cooling pool, which normally contains about 1500 tons of water, may be estimated 
as $220 \times 1.5 \times 10^9 = 3.3 \times 10^{11}$ Bq (on April 13) which is small 
compared to the radioactivity of fresh fission products contained in the nuclear reactor, 
which is the order of $10^{18}$ Bq (on X-day).   
If we assume that these I-131 isotopes detected in the Unit-4 cooling pool were produced 
by nuclear fissions reignited inside the pool after X-day, the upper bound of the total 
number of fissions occurred may be estimated as 
$N_f = 3.3 \times 10^{11} \times 2^{32/8} \times \tau_I / f_I = 1.3 \times 10^{20}$, 
which would have produced an extra heat  $\Delta U = N_f \times 200~{\rm  MeV} 
= 4.1 \times 10^9 $J.  This extra heat produced is equivalent to a 
latent heat absorbed by about 2 tons of boiling water, which is not significant if one
compares it to the decay heat produced daily during the period by the used fuel rods 
in the cooling pool and therefore may have been easily unnoticed.  
This, however, may not exclude a possibility that nuclear reaction had reignited on a larger 
scale and some extra fission products had escaped from the pool together with violently boiling 
water by sudden release of large amount of heat.}


\begin{figure}[t]
\begin{center}
\resizebox*{!}{6.6cm}{\includegraphics{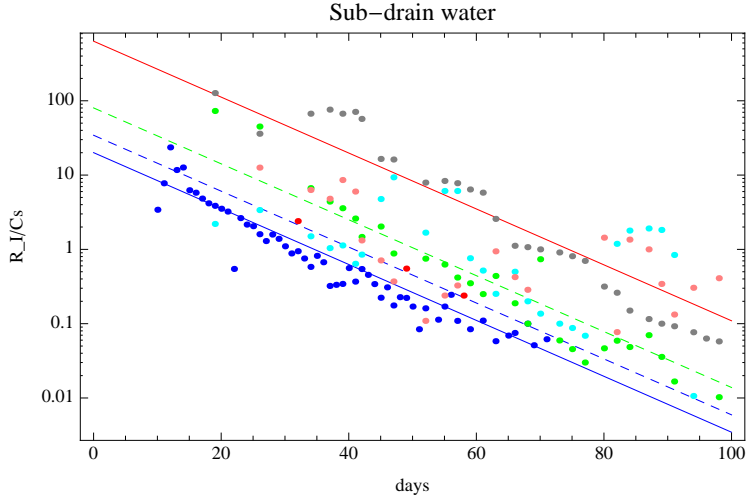}}
\end{center}
\caption{\label{drain} (color online)
Same plot as Fig.~1, now including the data of the water samples taken 
from the sub-drains near the Unit-1 (green), Unit-2 (gray), 
Unit-3 (cyan) and Unit-4 (pink) reactor buildings.  
Two more theoretical lines are shown as a  guide to the eye;  
they are computed from formula (\ref{R}) with $\Delta t = 3$ months (dashed green) and 
formula (\ref{Rmax}) with $t'_f = t_X$ (solid red), which gives an upper bound 
if the fission ended on X-day. 
}
\end{figure}

(iii) {\it Sub-drain water:}
The data from water samples taken from four sub-drains near the reactor 
buildings show even more puzzling features, as shown in Fig.~\ref{drain}.\cite{sub-drain}
In particular, the water samples from the sub-drain near the Unit-2 reactor 
building show an anomalously high radioactivity ratio\footnote{
The author thanks Gordon Baym for calling his attention to this point.} 
even greater than the upper bound given by (\ref{Rmax}) 
if the nuclear fission ends on X-day as indicated by the red solid line in the figure. 
If there is no strong chemical filtering effect in draining contaminated water 
from the reactor buildings, 
it would be difficult to understand the observed anomaly near the Unit-2 reactor 
without assuming that a significant amount of fission products were produced 
at least 10 -- 15 days after X-day.
We note that several step-wise sequential characteristic changes have been observed 
in the Unit-2 sub-drain data with largest change on May 11 (60 days from X-day).  
The causes of these changes are still unknown to the author, but it is tempting to speculate 
that they were caused by the mixing of the new fission products with the old ones 
released three times later from the remaining fuel rods in the Unit-2 reactor or from the 
water from other reactor site, as we shall see later. 

The data from the Unit-3 sub-drain before April 23 sit close to the decay line
which fits the sea water data, hence they may be understood as due to 
radiation from fission products produced before the X-day.
However, the data of the Unit-1 sub-drain and Unit-4 sub-drain give a high 
radioactivity ratio, even larger than that of the samples from the Unit-4 cooling 
pool and from the Unit-3 sub-drain. 
The data  therefore cannot be explained by the contamination of 
the old fission products which had existed in the spent-fuel rods 
in the Unit-4 cooling pool. 
The data occasionally show very different characteristics which are difficult to be 
understood unless there was considerable mixing of waters belonging to 
different sub-drains due to in-flow and out-flow of water through underground 
water channels. 
We note that the anomalously large iodine-cesium ratios were observed also in the sub-drain 
water in Unit-3 and Unit-4 reactor sites in June, and I-131 has been detected in the sub-drain 
water at Unit-1 and Unit-3 reactor sites briefly even as late as August 31, July 28, respectively.
It is difficult, however, to judge the significance of these data since there were considerably 
large fluctuations.

\begin{figure}[t]
\begin{center}
\resizebox*{!}{4.7cm}{\includegraphics{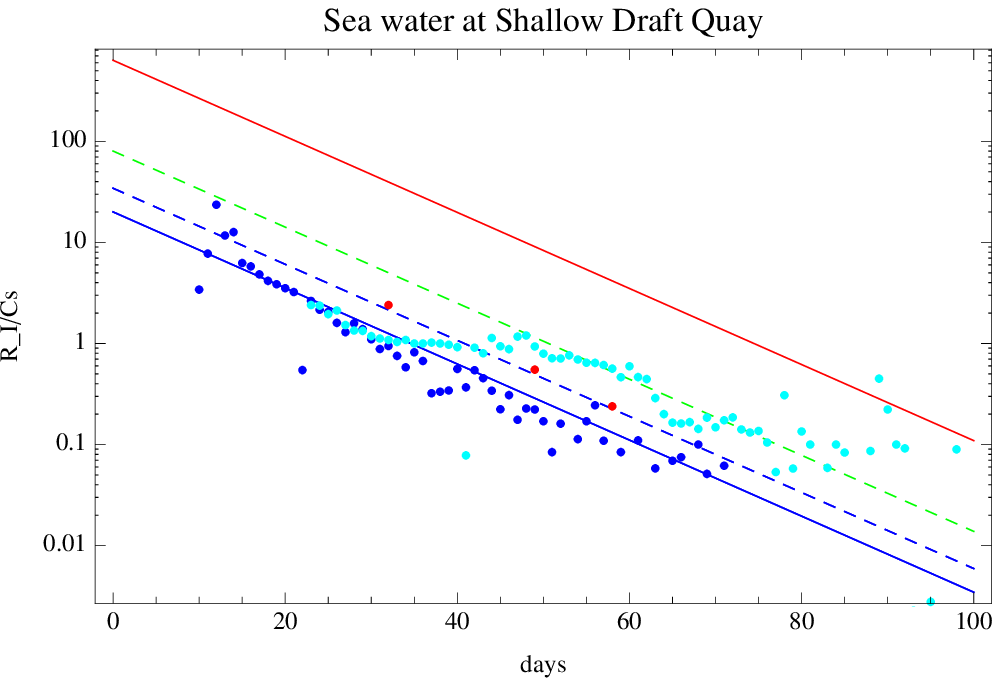}
\includegraphics{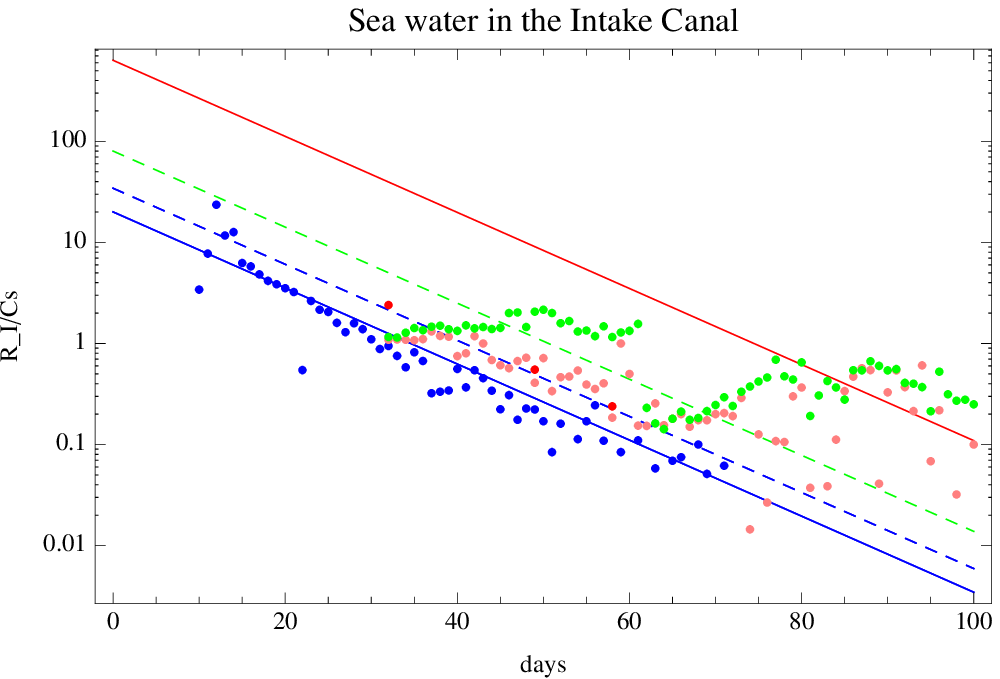}}
\end{center}
\caption{\label{canal} (color online)
Same plot as Fig.~1, with the data samples from  the shallow draft quay (cyan) and
at north (green) and south (pink) sides of the Intake Canal of Unit-1 -- 4 reactors.
}
\end{figure}

\begin{figure}[t]
\begin{center}
\resizebox*{!}{4.7cm}{\includegraphics{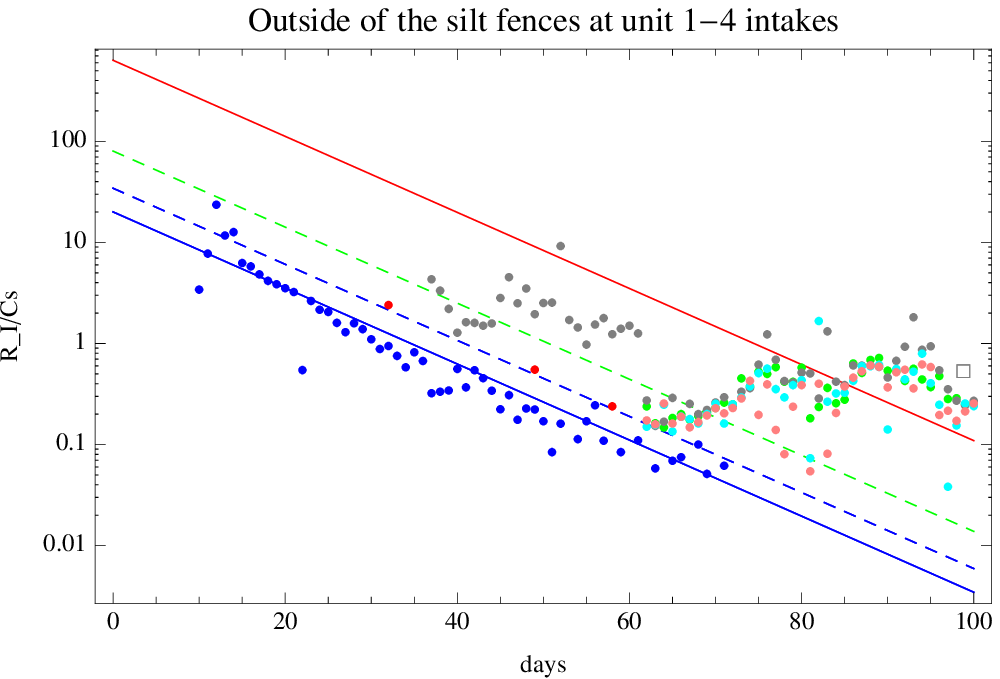}
\includegraphics{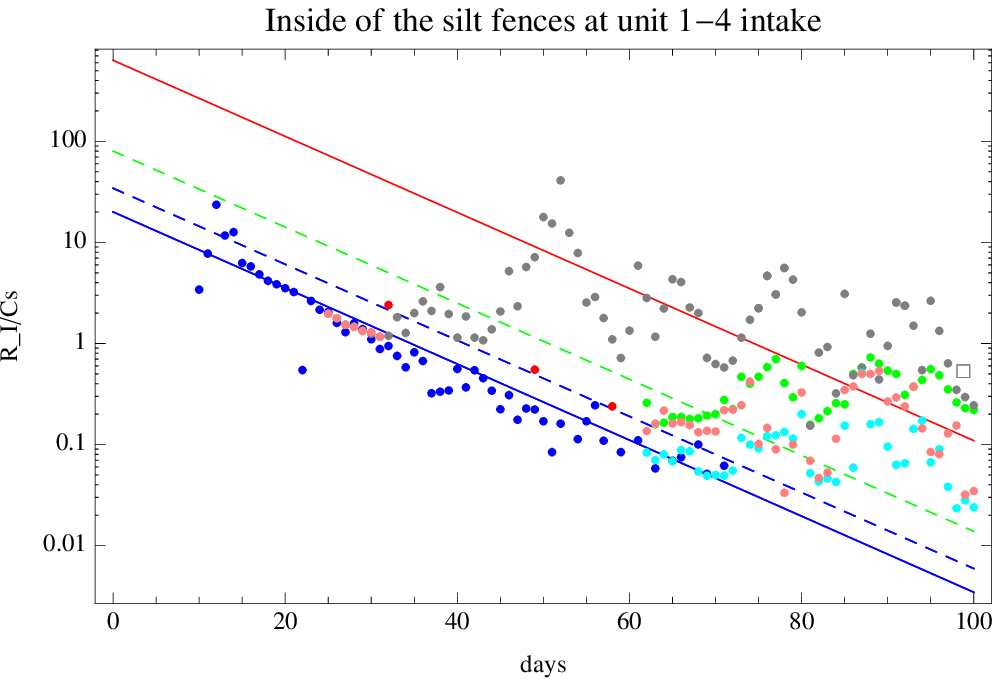}}
\end{center}
\caption{\label{screens} (color online)
Same plot as Fig.~1, with the data samples from the outside (left panel) and inside (right panel) 
of the silt fences at bar screens of Unit-1 (green), Unit-2 (gray), Unit-3 (cyan) and  Unit-4 (pink) reactor 
sites.  The open square is the ratio of I-131 and Cs-137 measured for the highly humid air (99.9\%) sampled 
inside the Unit-2 reactor building.  
}
\end{figure}

(iv) {\it Intake Canal water:}
We show a few more plots of the data available for sea water samples taken near the Intake Canal of 
the Unit-1 -- 4 reactor sites that may fill the gap between the previous sea water data (Fig.~1) and 
the sub-drain water data (Fig.~2).\footnote{
The author thanks Ryogo Hayano for calling his attention to the 
Intake Canal data.}
We note that a leakage of contaminated water from the water-intake of Unit-2 reactor had been reported 
on April 2nd, and this prompted to monitor the radioactive isotopes contained in the contaminated 
water near the intake of Unit-2 reactor.\cite{u2leakage}  
The left panel of Fig.~3 corresponds to the sea water at the Shallow Draft Quay, located just outside 
of the canal, while the right panel is for the samples taken at the north (green) and south (pink) 
sides of the Intake Canal: the north monitoring post is located near the Unit-1 reactor intake, 
while the south post is near the Unit-4 reactor intake.    
Although the available data is limited, all these ratios seem to have stayed very close to the values of 
the sea water data taken at the southern monitoring post 330~m away from the canal until April 10 
(30 days from X-day) and then started to deviate; 
the deviation becomes bigger at the north monitoring post than at the south, indicating that it is caused 
by the change of character of the sea water due to mixing with new contaminated water leaking out 
from the Unit-2 reactor.  
A sudden drop of the ratio has been observed on May 10  (60 days after X day) in these plots.   
This change may be caused by the leakage of contaminated water found in the pit of Unit-3 
reactor sites; the iodine-cesium ratio may have been small in the water leaking out from the Unit-3 
reactor and hence considerably diluted the ratio of pre-existing water in the canal with large iodine-cesium 
ratio.\cite{u3leakage}
After May 11 when the leakage from the intake of Unit-3 reactor was stopped, the ratios started to increase 
again and the Intake Canal data exceed the upper bound (red line) allowable, had the nuclear reaction 
terminated on the X day. 
We would also like to note that the iodine-cesium ratio in the contaminated water leaked from 
the Unit-3 reactor site is very close to the ratio found in the southern monitoring post 330~m south of these 
troubled reactors.  
This may not be just a coincidence.

To locate the origin of the increase of the ratios, we plot in Fig.~4 the data taken at the bar screens of the 
individual water intakes for Unit-1 -- 4  reactors: 
the left (right) panel is a plot of the data taken outside (inside) of the silt fences which have been 
installed at the bar screen of the intake of the Unit-2 reactor site since April 18, 
and at the bar screens of the other three reactor sites since May 12. 
The radioactivity ratios of iodine-131 and Cs-137 for the water outside of the fences show very 
similar behavior, with striking similarity to the north Intake Canal data (the green dots in the 
right panel of Fig. 3). 
On the other hand, the water sample taken inside the silt fences shows clear difference of 
characteristics from the water coming from individual reactor sites: the water coming from the pit of 
the Unit-2 reactor site has in particular an anomalously large iodine-cesium ratio.
This is qualitatively similar to what we have observed for the sub-drain water, while the ratio
fluctuates more rapidly in the water leaking out of the pit of Unit-2 reactor site. 
It is apparent that the contaminated water leaking out of the Unit-2 reactor site is responsible for 
the over-all changes of the iodine-cesium ratios of the other parts of the water in the Intake Canal.
It is interesting to note that the air sample taken from a very humid (99.9\%) air inside the Unit-2 
reactor building has a very high iodine-cesium ratio of 0.51 even 99 days after X day,
close to the value measured for the water coming out of the Unit-2 pit. \cite{air}\\

\noindent
4. {\it Conclusions: }\hspace{7mm}
In these notes we have estimated the radioactivity ratios of iodine isotope I-131 and 
cesium isotope Cs-137 accumulated in the nuclear reactor and compared the results with the data
of the Fukushima nuclear reactor accidents released by TEPCO for the first 100 days after the 
outbreak of the accident.  
Although there are still many uncertainties in connecting our estimates of the ratio in the nuclear 
reactor to the ratios measured in the water sampled outside of of the reactors, as mentioned in the 
very beginning of these notes, the results are rather provocative. 

The ratio of the measured radioactivity of  I-131 and Cs-137 may be used to extract useful information 
about when these fission products were produced in the nuclear reactor complex of the Fukushima 
Dai-ichi plant.   
The sea water data taken at southern monitoring post are consistent with the expected radioactivity 
by the fission products which had been produced before the earthquake and perhaps 
released by the hydrogen explosions of Unit-1 and -3 reactor buildings.  
The data of the water samples from the Unit-4 cooling pool and from 
areas near the Unit-2 reactor, however, show an anomaly which may indicate, 
if the data are correct, that some of these fission products were produced by 
chain nuclear reactions reignited after the earthquake.  

Some of these results may well be affected by chemical, thermodynamic and/or transport effects 
which are not taken into account in the present analysis. 
Estimates of these effects are rather difficult problems which would require detailed knowledge of 
the course of the accident in addition to various engineering problems related to the reactor design.   
Hopefully, more direct information concerning the interior conditions of the troubled reactors and 
the used fuel rods in the cooling pool will eventually become available and will clarify what actually 
had happened at the Fukushima Dai-Ichi reactor complex.  
 
\vskip 10pt
The author thanks Professors Gordon Baym, Munetake Ichimura and Koichi Yazaki 
for their interest in this work and for helpful communications.  
In particular, he is grateful to Professor Ichimura for patiently elucidating 
some of the data released by TEPCO and to Professor Baym for carefully reading 
these notes. 
He is grateful to Professor Ryogo Hayano for sharing his insights on many related 
aspects of the data for the Fukushima nuclear accident and for calling the author's 
attention to the intake canal data, and to Professor Mamoru Shimoi for information 
regarding the water-solubility of iodine and cesium elements.

\end{document}